\documentstyle[preprint,aps]{revtex}
\widetext
\draft
\tighten

\begin{document}

\title{Polar optical phonons at $ GaAs|Al_{(x)}Ga_{(1-x)}As$  interfaces:
 influence of the concentration of $ Al$}

\bigskip

\author{{\bf A. E. Chubykalo\thanks{Corresponding author. {\bf FAX}: (52)
492 4-13-14; {\bf e-mail}: andrew@cantera.reduaz.mx}, D.  A.
Contreras-Solorio and M.  E.  Mora-Ramos}}

\address {Escuela de F\'{\i}sica, Universidad Aut\'onoma de Zacatecas \\
Antonio Doval\'{\i} Jaime\, s/n, Zacatecas 98068, ZAC., M\'exico}

\maketitle

\baselineskip 7mm

\begin{abstract}
We study long-wavelength polar optical modes at semiconductor interfaces of
$GaAs|Al_{(x)}Ga_{(1-x)}As$ and take into account influence
of the $Al$ concentration. We have considered two cases  in
which the  interface is kept at unfixed and fixed electrostatic potential.
The spectrum of excitation then shows  existence of localized and resonant
modes to depend on the concentration of $Al$. For the case of a fixed
electrostatic potential, a splitting of the resonant mode near the
transverse threshold is found for certain ranges of the concentration $x$.
The existence of these modes has been obtained by computing the spectral
strength through the Surface Green Function Matching (SGFM) method.

\end{abstract}


\newpage
\section{Introduction}

During the last few years, there has been a growing interest in the study
of the phonon properties in semiconducting nanostructures, mainly those
concerning to long wavelength polar optical oscillations, Different
approaches have been present in order to give an appropriate
phenomenological description of polar optical phonons in the long
wavelength limit [1-4]. In that scenario, there are reported some results
 in the cases of layered structures [5-9], quantum wires [10-12] and
quantum dots [13]. In a recent paper [9] the spectrum of polar optical
modes for $GaAs$ surfaces and $GaAs|Al_{(0.9)}Ga_{(0.1)}As$ interfaces was
studied by using the theory developed in [2,5,6] and later applied to the
study of those modes in $GaAs$-based double heterostructures, with
remarkable agreement with microscopic {\it ab-initio} calculations
[14,15].

In the paper mentioned above [9], for free surfaces localized and resonant
modes were found. When assuming a metallized surface, thus modifying the
electrostatic boundary conditions, the {\it localized} mode is almost
unaffected while the {\it resonant} mode disappears,  thus evidencing a
strong electrostatic character.

In the case of the interface, where only {\it resonant} modes exist,
these modes are also affected when the interface is metallized, thus also
manifesting its strong electrostatic character. This explains why the
dielectric model (see e.g.  [16]) successfully accounts for the interface
modes found in quantum well systems.  In quantum  wells the modes having
predominantly electrical character can be identified with the {\it
interface} modes [17]. As we have seen for just one  interface one obtains
only one such mode. When the symmetrical quantum  well is formed the two
degenerate modes corresponding to the two interfaces split into the two
interfaces modes of opposite parities. The identification of the {\it
interface} modes as {\it predominantly} electrostatic is also relevant for
the theory of electron-phonon interaction in heterostructures when polar
optical modes are involved.  Since the measure of this interactions is
$e\varphi$ the analysis suggests that this should be dominated by the role
of the {\it interface} modes [20].  In our work we shall concentrate
mainly on influence of concentration $x$ of $Al$ on existence and
 ``behavior" of polar optical {\it localized} and {\it resonant} modes at
 the $GaAs|Al_{(x)}Ga_{(1-x)}As$ interface.

\section{Theoretical and practical aspects of the model}

Let us consider a medium where ${\bf u}({\bf r},t)$ is the
displacement and $\varphi({\bf r},t)$ is the scalar potential
associated with the electric field ${\bf E} = - \nabla\varphi$.
We have a mechanical equation of motion for ${\bf u}$ which is of the
form [2]
\begin{eqnarray}
 \rho(\omega^{2}-\omega^{2}_{TO}){\bf u} + {\mbox{\boldmath
 $\nabla$}}\cdot{\mbox{\boldmath $ \tau$}} - \alpha \nabla\varphi = 0 ;
 \nonumber \\ \alpha^{2} = \omega^{2}_{TO}\rho(\epsilon_{0} -
 \epsilon_{\infty})/4\pi.
 \end{eqnarray}
The harmonic oscillator part is contained in the first term. The
second term has the nature of a dispersive mechanical term and for
an isotropic medium is of the form
\begin{equation}
\tau_{ij} = - \rho(\beta_{L}^{2} - 2\beta_{T}^{2})\delta_{ij}
{\mbox{\boldmath $\nabla$}}\cdot{\bf u} - \rho
\beta_{T}^{2}(\nabla_{i}u_{j}+\nabla_{j}u_{i}),
\end{equation}
where $\nabla_{i}=\partial/\partial x_{i}$ and $\beta_{L}$ and
$\beta_{T}$ are adjustable parameters. The sign of (2) is
opposite to the usual one for acoustic waves on account of the
negative dispersion of the optical modes. The third term measures
the effect of the coupling between
the ${\bf u}$ and $\varphi$ fields on the equation of motion
for ${\bf u}$. Simultaneously we have a Poisson equation for
$\varphi$ which reads
\begin{equation}
 \nabla^{2}\varphi = 4\pi \gamma{\mbox{\boldmath $\nabla$}}\cdot{\bf u} ;
 \;\;\;\; \gamma=\alpha/\epsilon_{\infty}.
 \end{equation}
 The physical
meaning of the above equation is that $\varphi$ is created by the
polarization charge $\rho_{e} = \nabla\cdot {\bf P}$ of the polarization
field given by:  \begin{equation} {\bf P} = \alpha {\bf u} +
\frac{\epsilon_{\infty}-1}{4\pi} {\bf E} = \alpha {\bf u} -
\frac{\epsilon_{\infty}-1}{4\pi} \nabla\varphi \end{equation} The term on
the r.h.s. of (3) measures the effect of the coupling between the
$\varphi$ and ${\bf u}$ fields on the field equation for $\varphi$. Of
course we are working in the quasistatic limit ($c\rightarrow \infty$)
which is fully justified for the situation under study.

For an isotropic bulk homogeneous medium it is
possible to obtain independent equations for ${\bf u}_{L}$ and
${\bf u}_{T}$ and this yields at once the longitudinal and
transverse modes with dispersion relations
\begin{equation}
\omega^{2}_{L}=\omega_{LO}^{2} - \beta_{L}^{2}k^{2}; \;\;
\omega^{2}_{T}=\omega_{TO}^{2} - \beta_{T}^{2}k^{2} ,
\end{equation}
where ${\bf k}$ is the 3D wavevector and
$\omega_{LO}^{2}=\omega_{TO}^{2}(\epsilon_{0}/\epsilon_{\infty})$.
However, our concern is to study the matching of different media
at an interface.

Let  discuss one interface which we take as the plane
$z=0$. We first Fourier transform in the 2D plane of the
interface, so that the $\omega$-dependent vibration amplitudes are of the
form
\begin{equation}
 {\bf u}({\mbox{\boldmath $\rho$}},z) = \exp(i{\mbox{\boldmath
$\kappa$}}\cdot{\mbox{\boldmath $\rho$}}){\bf u}(z), \end{equation} where
${\mbox{\boldmath $\kappa$}},{\mbox{\boldmath $\rho$}}$ are 2D vectors
(wavevector and position) respectively.

We proceed likewise for $\varphi$ and concentrate on ${\bf u}(z)$
and $\varphi(z)$. These are $(\omega,{\mbox{\boldmath $\kappa$}})$-dependent
quantities for which, after 2D Fourier transform, we have
 $(\omega,{\mbox{\boldmath $\kappa$}})$-dependent differential equations
in the independent variable $z$. We stress that in general $\bf u$
consists of ${\bf u}_{L}$ and ${\bf u}_{T}$ (see, e.g., [8]).

The matching boundary conditions were obtained in [2] and
they are, in the variable $z$,
\begin{eqnarray}
u_{j}(+0)=u_{j}(-0) \nonumber \\
\varphi(+0) = \varphi(-0) \nonumber \\
\tau_{zj}(+0)=\tau_{zj}(-0) \nonumber \\
\epsilon_{\infty} \frac{d\varphi}{dz} - 4\pi\alpha u_{z}
= {\rm continuous}
\end{eqnarray}

On evaluating these expressions at $z=\pm 0$ we must account for
the different values of $\beta_{L}$, $\beta_{T}$,
$\epsilon_{\infty}$ and $\alpha$ on both sides of the interface.

We have four amplitudes ($u_{x},u_{y},u_{z},\varphi$), four coupled
second order linear differential equations and eight matching
boundary conditions. As in the study of piezoelectric surface or
interface waves it proves convenient to define a tetrafield
\begin{equation}
{\bf F}\;\equiv\; \left[\begin{array}{c}
{\bf F}_{M} \\ F_{E}
\end{array}\right] \;\equiv\; \left[\begin{array}{c}
{\bf u} \\ \varphi
\end{array}\right]
\end{equation}
which has mechanical and electrical components and to condense
the system (1), (3) in the form
\begin{equation}
{\bf L}\cdot{\bf F} = 0 ,
\end{equation}
where ${\bf L}$ is a 4$\times$4 differential matrix which can be
readily written down explicitly. Upon 2D Fourier transform ${\bf
L}$ depends on ${\mbox{\boldmath $\kappa$}}$ ($\omega$-dependence understood
everywhere) and contains the differential operator $d/dz$. A
convenient technique for solving this problem is the Surface
Green Function Matching (SGFM) method [19], which, for the case we are
interested on, was applied in [9]. In the SGFM method, the key element is
the surface projection of the Green function of the system ${\cal G}_s$.
The eingenvalues $\omega({\mbox{\boldmath $\kappa$}})$ of the problem can
be obtained from an procedure in which we calculate from ${\cal G}_s$ the
${\mbox{\boldmath $\kappa$}}$-resolved local density of states at $z=0$
[19]:
 \begin{equation} {\cal N}_s(\omega,{\mbox{\boldmath $\kappa$}})=
 -\frac{1}{\pi}\lim_{\varepsilon\to\infty}\;{\tt Im}\;{\tt Sp}\;{\cal G}_s
(\omega+i\varepsilon,{\mbox{\boldmath $\kappa$}}).
  \end{equation}
 The
eingenvalues $\omega({\mbox{\boldmath $\kappa$}})$ are then the
 frequencies at which the peaks in the density of states ${\cal N}_s$
  appear. This procedure is very convenient, especially when resonant
  modes can exist. Such resonant modes will appear as Lorentzian peaks
   superimposed on the continuum of bulk scattering states in the density
   of states.

Since we have studied $GaAs|Al_{(x)}Ga_{(1-x)}As$ systems, some
words about the fitting procedure employed to estimate the input
parameters are in order. The mass density and the background
dielectric constants were obtained from a linear interpolation
of the value for the pure materials - $AlAs$ and $GaAs$ -
according to the formulae [20]:
\begin{eqnarray}
\rho(x) & = & 5.36 - 1.60 \; x \; ; \; {\rm (c.g.s.)} \nonumber
\\
\epsilon_{0}(x) & = &13.18 - 3.12 \; x \; ; \; {\rm (e.s.u.)} \nonumber
\\
\epsilon_{\infty}(x) & = &10.89 - 2.73 \; x \; ; \; {\rm (e.s.u.)}
\end{eqnarray}
However, this type of interpolation would not work for $\omega_{LO}$,
$\omega_{TO}$, $\beta_{L}$ and $\beta_{T}$. It has been
strongly argued on the basis of experimental evidence [20] that
the ternary compound $Al_{x}Ga_{(1-x)}As$ can be described in
the {\em two-mode model}. We shall adopt this viewpoint.  It then follows
that if we study the matching to $GaAs$ we must assign to the ternary
alloy the values of the frequencies for the $LO$ and $TO$ modes found
experimentally for the $GaAs$ like modes in this alloy [21]
\begin{eqnarray}
\omega_{LO,GaAs}(x) & = & 292.37 - 52.83 \; x \; + \; 14.44 \;
x^{2} \; ; \nonumber \\
\omega_{TO,GaAs}(x) & = & 268.50 -  5.16 \; x \; - \;  9.36 \;
x^{2}
\end{eqnarray}
Here and henceforth $\omega$ is always given in
cm$^{-1}$.

The $\beta_{L}$ and $\beta_{T}$ parameters are not usually
reported in the literature and they are not known for the
different types of modes in alloys. We have estimated their
values for the pure materials $(x=0, x=1)$ from the experimental
curves of Ref. [21].

We have also made the following assumption: For very low (high)
concentrations of $Al$, that is for $x\approx 0(x\approx 1)$, we
take dispersion laws with $\beta=0$ for the $AlAs$($GaAs$)-like
modes. This assumption relies on the fact that for these
situations the atoms in question are isolated and their phonon
branches must be flat. For a given concentration $x$, we perform
a linear interpolation between the values for $x=0$ and $x=1$.
With $\omega$ in cm$^{-1}$, $\beta$ is dimensionless and we obtain
\begin{eqnarray}
\beta^{2}_{L,GaAs}(x)& = & 2.91 \; (1-x) \; 10^{-12} , \nonumber \\
\beta^{2}_{T,GaAs}(x)& = & 3.12 \; (1-x) \; 10^{-12}
\end{eqnarray}

\section{Results and conclusions}
 Now we are going to discuss the results obtained by the model considered
above.  A small imaginary part,  equal  $10^{-6}{\tt cm}^{-1}$, was added
to $\omega$, so one can ascribe the local modes to very narrow peaks,
which show in the calculations instead of the ideal $\delta$-functions.
 Repeating the calculations for different values of $x$ in
  $GaAs|Al_{(x)}Ga_{(1-x)}As$  we have observed the concentration of $Al$
   to have influence on ``behavior" of localized and resonant modes
   obtained in [9].

 We are going to start with the study of the $GaAs|Al_{(x)}Ga_{(1-x)}As$
non-metallized interface. In our figures we present the spectral strength
for value of $\kappa=4\times 10^6{\tt cm}^{-1}$.
It is seen that no localized mode is present, while the peak
corresponding to the resonant mode is present ($\omega\simeq 282.6 {\tt
cm}^{-1}$). In Fig.1 we can see that the frequency of the resonant mode
slowly increases with decreasing $x$ and the peak, at last, disappears for
values below $x\simeq 0.5$. At the same time, from Fig.2 it can be seen
that the position of the localized mode in the transverse threshold moves,
as $x\rightarrow 0$, toward a peak corresponding to the frequency of the
transverse oscillations of $GaAs$ bulk crystal. A similar behavior can be
observed for the longitudinal one, thus we omit to show it.

We shall pass now to study  the case when the interface is metallized.
Here, a very interesting effect can be seen (Fig.3). As it was previously
reported [9], in this case the resonant mode moves toward the transverse
threshold and is very close to it. But when studying the dependence of its
position with $Al$ concentration  it appears that, for a very narrow range
$0.5292\geq x \geq 0.5308$ (with $\kappa =4\times 10^6 {\tt cm}^{-1}$),
the corresponding peak in the LDOS splits into two very well resolved
peaks located on both sides of the transverse threshold. When the
concentration is raised, the left peak disappears (Figs.4a,4b)  and the
remaining one locates near $\omega_{TO}$ as was obtained in [9]. We do not
know any previous experimental or theoretical report on that kind of
splitting, which can also be seen for other values of $\kappa$, but with a
slightly different range for $x$ (Fig.5). The present calculation does
not allow us to give a precise explanation for such an effect. We assume
that, regarding the strong electrostatic character of this mode, for those
values of $x$, the potential energy at the interface modifies in a way
that some kind of ``potential well" is present, leading to a situation
analogous to the case of a quantum well, where the two degenerate modes
corresponding to the two interfaces split into the two interface modes of
opposite parities. Of course, here we do not have two interfaces, but only
one, and only one of those modes splits. The ``confinement" for the
oscillations would come from the shape of the potential energy
distribution at the interface.  Nevertheless, the exact  explanation can
only be obtained by means of a detailed analysis of that distribution.
According to the results we are presenting here it seems that the study of
this problem deserves some attention and will be considered elsewhere.

Recently, a new envelope-function theory for phonons in heterostructures
[4] shows interesting results for the $InAs|GaSb$ system. Even when it
appears that the study of multiple-interfaces heterostructures is more
attractive looking forward possible applications, the consideration of a
single one still gives the possibility of a deeper understanding of the
effect of interfaces on the long-wavelength oscillations, so it would be
desirable to study, for instance, the presence of the mechanical interface
modes, which occurred in those systems.

\bigskip

{\large Acknowledgments}

We are grateful to Dr. S. Vlaev and Prof. V.R. Velasco for many
stimulating discussions.  Authors are indebted for  financial support  to
the Zacatecas University, M\'exico. One of the authors (D.A.C.S.) is
indebted for financial support of EC (CI1$^*$-T94-006) and CONACyT
(3148-E9307).

\newpage

\section{Figure Captions}

{\bf Fig.1}: LDOS for the resonant mode in a
non-metallized interface for $x$ from $0.9$ to $0.5$. Curves, from {\it
left} to {\it right}, correspond to decreasing values of $Al$
concentration.

\bigskip
\bigskip
\bigskip

{\bf Fig.2}: LDOS for the localized transverse-like
oscillation mode in a non-metallized interface for $x$ from $0.2$ to 0.
Curves, from {\it left} to {\it right}, correspond to decreasing values of
$Al$ concentration.

\bigskip
\bigskip
\bigskip

{\bf Fig.3}: LDOS showing the splitting of the resonant mode at the
transverse threshold in a metallized $GaAs|Al_{(x)}Ga_{(1-x)}As$
interface for $\kappa =4\times 10^6{\tt cm}^{-1}$.

\bigskip
\bigskip
\bigskip

{\bf Fig.4(a)}: LDOS for a metallized interface. It is seen how one of
the peaks (the left one in Fig.3) resulting from the splitting
decreases its height with increasing $x$.

\bigskip
\bigskip
\bigskip

{\bf Fig.4(b)}: The continuation of LDOS from Fig.3 for higher values
of $x$ showing the disappearance of the left peak  ($\kappa=4\times
10^6{\tt cm}^{-1}$).

\bigskip
\bigskip
\bigskip

{\bf Fig.5}: Splitting in the LDOS for a metallized
$GaAs|Al_{(x)}Ga_{(1-x)}As$ interface for  $\kappa =2\times 10^6{\tt
cm}^{-1}$.


\begin{references}

\bibitem{1} B.K. Ridley and M. Babiker, Phys. Rev. B, {\bf 43}, 9096 (1991).
\bibitem{2} C.  Trallero-Giner, F. Garc\'{\i}a-Moliner,  V.R. Velasco and
M. Cardona, Phys. Rev. B, {\bf 45}, 11944 (1992).
\bibitem{3} K.J. Nash, Phys. Rev. B,  {\bf 46}, 7723 (1992).
\bibitem{4} B.A. Forreman, Phys. Rev. B, {\bf 52}, 12260 (1995).
\bibitem{5} F. Comas and C. Trallero-Giner, Physica B, {\bf 192}, 394
(1993).
\bibitem{6} R.  P\'erez-Alvarez, F.  Garc\'{\i}a-Moliner, V.R.
Velasco and C. Trallero-Giner, J. Phys.: Condens. Mat., {\bf 5}, 5389 (1993).
\bibitem{7} R. P\'erez-Alvarez, F. Garc\'{\i}a-Moliner, V.R.
Velasco and C. Trallero-Giner, Phys. Rev. B, {\bf 48}, 5672 (1993).
\bibitem{8} F. Comas, R. Perez-Alvarez, C. Trallero-Giner and M. Cardona,
Superlatt. and Microstruc., {\bf 14}, 95 (1993).
\bibitem{9} A. Chubykalo, V.R. Velasco and F. Garc\'{\i}a-Moliner,
Surf. Sci., {\bf  319}, 184 (1994).
\bibitem{10} Sh-F. Ren and Y-Ch. Chang, Phys. Rev. B, {\bf 43}, 11857
(1991).
\bibitem{11} F. Rossi, L. Rota, C. Bungaro, P. Lugli, and E. Molinari,
Phys. Rev. B. {\bf 47}, 1695 (1993).
\bibitem{12} F. Comas, C. Trallero-Giner and A. Cantarero, Phys. Rev. B,
{\bf 47}, 7602 (1993).
\bibitem{13} E. Roca, C. Trallero-Giner and M. Cardona, Phys. Rev. B, {\bf
49}, 13704 (1994).
\bibitem{14} H. R\"ucker, E. Molinari and P. Lugli, Phys. Rev. B, {\bf
44}, 3463 (1991).
\bibitem{15} H. R\"ucker, E. Molinari and P. Lugli, Phys. Rev. B, {\bf
45}, 6747 (1992).
\bibitem{16} X. Zianni, P.N. Butcher and J. Dharssi, J. Phys.: Condens.
Mat., {\bf 4}, L77 (1992).
\bibitem{17} F.Garc\'{\i}a-Moliner, {\it
Phonons in Semiconductors}, (Kluwer, Dordrecht, 1993).
\bibitem{18} M. E. Mora-Ramos, (Ph.D Thesis, Univ. of Havana, 1995).
\bibitem{19} F. Garc\'{\i}a-Moliner
and  V.R. Velasco, {\it Theory of Single and Multiple Interfaces}, (World
Scientific, Singapore, 1992).
\bibitem{20} S. Adachi, J. Appl. Phys., {\bf
58}, R1 (1985).
\bibitem{21} Z.P. Wang, D.S. Jiang and K. Ploog, Solid St.
Commun., {\bf 65}, 661 (1988).

\end{references}
\end{document}